\documentclass{article}[12pt]
\usepackage[utf8]{inputenc}
\usepackage{fullpage}
\usepackage{amsmath}
\usepackage{amsthm}
\usepackage{amssymb}
\usepackage{natbib}
\usepackage{xcolor}
\usepackage{booktabs, graphicx}
\usepackage{pdflscape} 

\title{Pseudo P-values for Assessing Covariate Balance in a Finite Study Population with Application to the California Sugar Sweetened Beverage Tax Study}
\author{Bing Han$^1$, Margo A. Sidell$^1$}
\date{\today}
\begin{document}
\maketitle
1. Southern California Kaiser Permanente

\begin{abstract}
    Assessing covariate balance (CB) is a common practice in various types of evaluation studies.  Two-sample descriptive statistics, such as the standardized mean difference, have been widely applied in the scientific literature to assess the goodness of CB.  Studies in health policy, health services research, built and social environment research, and many other fields often involve a finite number of units that may be subject to different treatment levels.  Our case study, the California Sugar Sweetened Beverage (SSB) Tax Study, include 332 study cities in the state of California, among which individual cities may elect to levy a city-wide excise tax on SSB sales. Evaluating the balance of covariates between study cities with and without the tax policy is essential for assessing the effects of the policy on  health outcomes of interest. In this paper, we introduce the novel concepts of the pseudo p-value and the standardized pseudo p-value, which are descriptive statistics to assess the overall goodness of CB between study arms in a finite study population.  While not meant as a hypothesis test, the pseudo p-values bear superficial similarity to the classic p-value, which makes them easy to apply and interpret in applications.  We discuss some theoretical properties of the pseudo p-values and present an algorithm to calculate them.  We report a numerical simulation study to demonstrate their performance. We apply the pseudo p-values to the California SSB Tax study to assess the balance of city-level characteristics between the two study arms. 

\end{abstract}
\newpage
\section{Introduction}
Covariate balance (CB) is a critical notion in causal inference for various non-experimental studies
\citep{rosenbaum1983central, ali2015reporting, pattanayak2011propensity, stuart2010matching}.  
Conceptually, the goodness of CB refers to the similarity in the joint distribution of all covariates between  study arms.  Practically, the goodness of CB is usually assessed by some two-sample descriptive statistics, such as the standardized mean difference or the p-value from a two-sample test \citep{stuart2010matching}. 
 Various methods based on the propensity score and its variants have been proposed in the literature to find study samples that have a satisfactory level or the best possible level of CB \citep{zhang2019balance, cannas2019comparison, stuart2010matching, austin2019assessing, pattanayak2011propensity, imai2014covariate}.  
Alternative methods not relying on propensity scores have also been developed offering advantages such as faster computation, better levels of CB, or improved efficiency in estimating the average treatment effect downstream \citep{pimentel2015large,
zhao2015minimal, yu2020matching, wang2021flame}.  

Theoretically, experimental studies such as randomized controlled trials (RCT) guarantee unconfoundedness between  covariates and study condition assignments through experiment design and randomization procedures.  However, the level of CB can still be assessed in an RCT for several reasons.  
First, the assessment of CB can be seen as detailed descriptive statistics of the study sample, which does not affect the experiment design. In the recent medical literature on RCT studies, it is common practice to report some kind of CB descriptive statistics such as standardized mean differences or p-values, e.g., \citet{lu2024perioperative, bekelmanadaptrct, oldenburg_azithrct}, and many more. 
Second, imperfect study executions, such as noncompliance or loss of follow-up, can compromise the ideal CB established by initial randomization. Therefore, checking CB becomes necessary in pragmatic trials, e.g., \citet{agniel2022synthetic, meredith2016impact}.  
Third, randomization may not guarantee that the selected sample is free of any large discrepancies, e.g., sample moments can differ substantially between arms in an ideal RCT. Sometimes assessment of CB may become a part of the study design, e.g., re-randomization to ensure or improve CB \citep{morgan2015rerandomization} and randomization within matched pairs \citep{greevy2004optimal}.  
However, there are serious concerns on allowing CB to play a role larger than simple descriptive statistics in RCT for both theoretical and practical reasons. It may distort the unconfoundedness of an ideal randomization procedure,  create artificial residual biases, or even lead to malpractice in applied statistics  \citep{mutz2019perils, imai2008misunderstandings}. 

While most existing methods focus on large-sample properties of CB, we note that the common practice of assessing CB through two-sample descriptive statistics can be highly restrictive for a relatively small sample and a finite study population. In Section 2.2, we provide examples to demonstrate the challenge. Throughout this paper, we use the term \textit{experimental unit} (EU) to refer to the smallest unit in an evaluation study that can receive a distinct study condition, regardless of whether the study is an experiment. 
Studies in health policy, health services research, built
and social environment research, and many other areas often have a small to medium-sized study population of EU, i.e., the collection of EU potentially accessible to a study.  In such a finite population, individual EU can be not only uniquely identifiable but also of special interest to various stakeholders. 

Our aim in this paper is to present a new global measure, called the \textit{pseudo p-value}, to fill in the gap in statistical methods to assess the overall goodness of CB between two study arms of a study sample drawn from a finite population. 
Our motivating application is the California Sugar-Sweetened Beverage (SSB) Tax Study \citep{SSB1, SSB2}. 
The SSB excise tax usually imposes a flat tax rate (e.g., 0.01 U.S. dollar per liquid ounce of SSB) to distributors, with the goal of raising additional government revenue dedicated to public health initiatives and mitigating sugar consumption at the population level. 
Various SSB excise tax policies are currently in effect in many non-U.S. nations. In the U.S., adoption of the SSB excise tax has been slow, and only at the regional or local level in recent years. 
The SSB excise tax has been a public topic of significant contention, attracting national attention since its introduction to residents of California in 2010s.  
Over time, four California cities (Albany, Berkeley, Oakland, and San Francisco) have voted to implement the SSB excise tax between 2015 and 2017.
The California SSB Tax Study is an important policy evaluation aimed at providing scientific evidence for the tax policy's impact on population health outcomes, such as obesity, body mass index, and incidence of new diabetes cases.  Results of the study can have state-wide and national implications for informing  future tax policy. 
The study is a natural experiment with two study arms, i.e., the presence or absence of a city-level SSB excise tax.  Besides the four cities currently having the SSB excise tax, other potential EU include incorporated California cities without the tax.  All study cities have health care services provided by Kaiser Permanente (KP). 
In this application, we have selected 40 unique comparison cities with 10 comparison cities matched to each of the four tax cities. 
The goodness of CB between the two study arms and between each tax city and its 10 comparison cities is the foundation for this evaluation study. We discuss more details of the application in Section 4. 

We also remark that a finite study population of EU is common in many research areas. In particular, cluster RCT studies often have access to a relatively small and finite number of clusters as EU.  For example, the COVER-HCW study randomized 28 hospitals and clinics
from two regional health care organization networks to a peer-to-peer support intervention versus the usual care mode \citep{coverhcw2024, dong2022protecting}.  \citet{cohen2017promoting} conducted a three-arm intervention study to promote physical activity in public space, where the study population of EU consisted of 80 staffed and equipped parks and recreation centers in the city of Los Angeles and the study sample included 50 parks. 
The Urban Gardens and Peer Nutritional Counseling Intervention on HIV trial consisted of only two eligible clinics in the Dominican Republic as EU for randomization \citep{derose2023preliminary}. 
The finite study population is not limited to cluster RCT. Many trials for orphan drugs 
have a small study population of patients and an even smaller study sample \citep{kanters2013systematic, mori2023efficacy, healey2023pexidartinib}. 
Many health policy studies are non-experimental, where an EU is often a government unit with jurisdictional authority. \citet{abadie2010synthetic} used individual states in the U.S. to evaluate the effect of a state-level cigarette sales tax policy.  \citet{klumpp2022covid} used countries as EU to assess the effect of nation-level mandatory measures during COVID-19.  
\citet{khomenko2021premature} used roughly 1,000 European cities as EU to study the impact of air pollution on mortality. There are many examples of a finite study population in a wide range of research areas. 
    
This paper is organized as follows. In Section 2, we introduce the notation and present the definition, properties, interpretation, and computation of the pseudo p-values.  In Section 3, we demonstrate the pseudo p-values by numerical simulation studies.  In Section 4, we apply the pseudo p-values to the California SSB Tax Study to assess balance of study city characteristics.   We conclude the paper by a short discussion in Section 5. The Appendix provides a sketch of proofs for the theoretical results in Section 2. Additional results from the simulation study and details of the study sample in the case study are included in the Supplement.

\section{Pseudo p-values for assessing covariate balance in a finite population}
\subsection{Notation and basic setting}
Adopting the classic notation in the survey sampling literature \citep{lohr2021sampling}, we denote a finite population of EU by $\mathcal U = \{1,\ldots,K\}$, with the population size $K$. 
Let $\mathbf X_i=(x_{i1},\ldots,x_{iJ})', i=1,\ldots,K$ denote $J-$dimension characteristics of all EU. Entries of $\mathbf X_i$ can further include functions of the raw information, e.g. $x_{i,3}= x_{i,1}^2$, $x_{i,4}=x_{i,1}x_{i,2}$. 
All $\mathbf X_i, i=1,\ldots,K$ are known. The population mean and variance for the $j$th characteristic are 
\begin{displaymath}\begin{split}
&\overline{x}_{j,U} =\frac{1}{K} \sum_{i=1}^K x_{i,j}, ~~~~
S_j^2 = \frac{1}{K-1}  \sum_{i=1}^K (x_{i,j} - \overline x_{j,U})^2.
\end{split}\end{displaymath}

Let $A$ be an observed sample drawn from $\mathcal U$ without replacement, i.e., $A$ is a fixed index set, $A \subset \mathcal U$.  Let $|A|$ be the sample size of $A$. The sample mean and standard deviation for the $j$th characteristic in sample $A$ are
\begin{displaymath}\begin{split}
&\overline{x}_{j, A} =\frac{1}{|A|} \sum_{i \in A} x_{i,j}, ~~~~
 s^2_{j,A} = \frac{1}{|A|-1}  \sum_{i\in A} (x_{i,j} - \overline x_{j,A})^2. \\
\end{split}\end{displaymath}
Given the observed sample $A$, $\overline{x}_{j, A}$ and $s^2_A$ are known constants. 

In a two-arm evaluation study, we define two observed and non-overlapping index sets $M$ and $N$ drawn from $\mathcal U$ without replacement.  The non-directional standardized difference in the sample mean (SMD) of the $j$th characteristic between samples $M, N$ is
\begin{align}\label{eqn:smd}
&\boldsymbol\Delta_j = \frac{ |\overline{x}_{j, M}  - \overline{x}_{j, N}|}{S_j}. 
\end{align}

It is worth noting that the SMD \eqref{eqn:smd} cab be replaced by many other descriptive statistics.  For example, a slightly different alternative to \eqref{eqn:smd} is to replace the denominator $S_j$ by the sample standard deviation, e.g., $s_{j,M}$, $s_{j,N}$, or $s_{j,M\cup N}$.  In the case of a binary characteristics, one can use the odds ratio,  relative risk, or Cohen's h between $M$ and $N$. A categorical characteristic with a finite number of categories can always be converted to a series of binary characteristics.   
The face value, but not the inference conclusion, of the classic p-value or the test statistic from a two-sample test suffices to serve in place of \eqref{eqn:smd} as well. Directions of the differences can also be preserved by removing the absolute value operation in \eqref{eqn:smd}.
Choices of these basic univariate measures do not affect our development. For simplicity and considering the popularity of the SMD in the literature, we use \eqref{eqn:smd} throughout this paper. 

Next, we also consider a sampling procedure, which can yield two random samples, or equivalently, two random index sets, denoted by $g, h$.  
The observed samples $M, N$ can be seen as one particular instance by some sampling procedure.   The SMD for the random samples is denoted as  
\begin{align}\label{eqn:smd2}
&\Delta_{j}(g,h) = \frac{ |\overline{x}_{j, g}  - \overline{x}_{j, h}|}{S_j}, j=1,\ldots, J.
\end{align}
$\Delta_{j}(g,h)$ is a random variable, whose randomness results from the random sampling of $g, h$. 

\subsection{A common covariate balance assessment procedure}
Let $\delta_j \ge 0 $ be a pre-specified cutoff on the SMD.  Namely, if $\boldsymbol\Delta_j\ge \delta_j$, then the $j$th covariate is assessed as ``imbalanced" between samples $M, N$. 
For simplicity in presentation, we assume a single cutoff $\delta=\delta_j, j =1,\ldots,J$.  Let $\mathbf R^\delta$ be the number of imbalanced characteristics between $M,N$, i.e., 
\begin{displaymath}\begin{split}
\mathbf R^\delta = \sum_{1\le j\le J}  I\{  \boldsymbol\Delta_j \ge \delta  \}. 
\end{split}\end{displaymath}
Given $M, N$, $\mathbf R^\delta$ is known.  A relatively large $\mathbf R^\delta$ usually leads to the conclusion that $M, N$ are ``imbalanced".  We formalize this common ad hoc procedure as follows. 

~\\

\begin{flushleft} 
\textit{An ad hoc procedure to check CB}
\end{flushleft}

\begin{enumerate}
    \item Specify a cutoff $\delta$, where conventional choices of $\delta$ is usually between .05 and .3.
\item  Specify a maximum number of imbalanced dimensions one is willing to accept, denoted as $r$, which is usually a small number such as 0 to 2, or a small proportion of $J$ when $J$ is large, e.g., $5\%$ of $J$. 
\item  Count the number of imbalanced dimensions $\mathbf R^\delta$. 
\item Declare the samples $M, N$ as balanced if $\mathbf R^\delta \le r$, and imbalanced otherwise. 
\end{enumerate}

~\\
Note that one or both of the cutoffs $r, \delta$ may not be fully specified in an application. There are also many minor variations of this procedure. 
This seemingly reasonable procedure can have great difficulties for a small sample and a finite population of EU. To illustrate the challenge, we consider two random samples $g, h$.  Let 
\begin{displaymath}    
R^\delta_{g,h} =\sum_{1\le j\le J}  I\{  |\Delta_j(g,h)| \ge \delta  \}
\end{displaymath}
denote the random number of imbalanced dimensions between $g,h$.  
Using the simple random sampling without replacement (SRS) procedure to draw $g, h$ and ignoring finite population corrections, we have 
\begin{displaymath} 
\begin{split}
&E(\overline{x}_{j, g}  - \overline{x}_{j, h}) =0,\\
&Var(\overline{x}_{j, g}  - \overline{x}_{j, h})
\approx (\frac{1}{n} + \frac{1}{m}) S_j^2 \\
&P( I\{ \Delta_j(g,h) \ge \delta\} ) 
= P( |\overline{x}_{j, g}  - \overline{x}_{j, h}|
\ge \delta S_j).\\
\end{split}
\end{displaymath}
If we use a normal distribution to approximate the last probability, it is 
\begin{displaymath}
P( I\{ \Delta_j(g,h) \ge \delta\} )
\approx 2-2\Phi(
(\frac{1}{n} + \frac{1}{m})^{-\frac{1}{2}}\delta ),
\end{displaymath}
where $\Phi()$ is the cumulative distribution function (CDF) or the standard normal distribution $N(0,1)$. 
Next, we use a binomial distribution with parameters $J, 2-2\Phi((\frac{1}{n} + \frac{1}{m})^{-\frac{1}{2}}\delta)$ to approximate the distribution of $R^\delta_{g,h}$. Then we can have a crude idea about the probability that an SRS sample is assessed as balanced, i.e., 
$P( R^\delta_{g,h} \le r)$. 
Table 1 provides some examples where the probability of declaring an SRS sample as balanced can be very low in  settings with a relatively small sample size.  For example, with an equal sample size of 40 per arm and $J=20$ covariates, the conventional cutoff $\delta=.2$ results in about 0.7\% SRS samples as balanced. Although a suitable algorithm can still find the few balanced samples, the collection of all so-called balanced samples may be an overly stringent study design.  The difficulty of the ad hoc procedure is eased when the sample size and the cutoff are relatively large: when the sample size is 100 per arm, the probability of declaring a balanced SRS sample increases quickly to 97.1\% with $\delta=.3, r=2, J=20$.   

Our case study, the California SSB Tax Study is a real data example to demonstrate the difficulty of the ad hoc procedure in practice. Similar sample size settings in Table 1 suggest that it is very difficult to find a balanced sample with $\delta=.2$ or $.3$.  Indeed that is the case in reality: by the ad hoc procedure, very few SRS samples can be assessed as balanced using conventional cutoffs, and the best matched sample using a weighted Euclidean distance metric among all study cities is anything but balanced by the ad hoc procedure. See Section 4 and Supplement for more details.

\subsection{The pseudo p-values}
Ideally, we would prefer to use an ideal sampling scheme to generate the study sample, e.g., an SRS or a stratified two-step sampling procedure given the sample size $|M|, |N|$.   Note that the ideal sampling strategy may or may not be able to generate the 
observed sample $M, N$.  Without loss of generality, let $g, h$ denote a random sample drawn from the ideal sampling scheme hereafter. Let the CDF of $R^\delta_{g, h}$ be denoted as
\begin{equation}
\begin{split}
F^{\dagger}_{\delta} (a) = P ( R^\delta_{g,h} \le a | \delta) = 
P ( \sum_j  I\{  |\Delta_j(g,h)| \ge \delta  \} \le a).
\end{split}\end{equation}
With respect to an ideal sampling scheme, we define a CB measure for the observed sample $M, N$, called as the pseudo p-value.  

\begin{flushleft}
    \textbf{Definition 1.} The pseudo p-value for the observed sample $M, N$ is
\end{flushleft}
\begin{equation}\label{eqn:p}
    \begin{split}
\mathbf p = \inf_\delta \big [ 1- F^\dagger_\delta (\mathbf R^\delta-1) \big ] = 1 - \sup_\delta F^\dagger_\delta (\mathbf R^\delta-1), 
\end{split}
\end{equation}
where an ideal sampling strategy gives $F^\dagger_\delta$, and $\delta>0$. 

The name of pseudo p-value is due to the superficial similarity between Definition 1 and the classic p-value in a hypothesis test.  If we ignore the limit in \eqref{eqn:p},  $1- F^\dagger_\delta (\mathbf R^{\delta}-1)$ is the tail probability that the ideal sampling scheme generates a random sample with no fewer imbalanced dimensions than the observed sample. This interpretation mimics one popular definition of the class p-value, i.e., the tail probability of observing more extreme values than the test statistic under the null hypothesis. 

Heuristically, a large pseudo p-value suggests that the observed sample $M, N$ has a good level of CB in relation to the ideal sampling strategy:  with a high probability a random sample drawn by an ideal sampling scheme has the same or more imbalanced dimensions than the observed sample $M, N$.  Vice versa, a small pseudo p-value suggests that the observed sample $M, N$ has a poor level of CB compared with the ideal sampling scheme.  Thus, if we are to pursue a better level of CB, a large pseudo p-value is preferred. This preference is similar to that of the classic p-value in a goodness-of-fit test, where larger p-values tend to suggest a satisfactory model fitting to data. 
Borrowing from the superficial similarity with the classic p-value, we tentatively suggest that conventional cutoffs of .01, .05, and .10 can be used to qualitatively assess CB by pseudo p-values. Namely, a pseudo p-value smaller than a customary cutoff such as .05 suggests that the observed sample is imbalanced, with the caveat that a larger cutoff (e.g., .10) is more stringent than a smaller cutoff (e.g., .01) to conclude a good level of CB.  
However, we must emphasize that the pseudo p-value is essentially a descriptive statistic and unrelated to hypothesis testing. In a real evaluation study, researchers know a priori whether the observed sample is generated by the ideal strategy. This fact is irrespective of the pseudo p-value.  

The pseudo p-value does not directly shed light on the size of the SMD. Instead, it aims to measure the goodness of CB between the observed sample and random samples from an ideal strategy.  The observed sample can have seemingly small values in many or all $\boldsymbol \Delta_j, j=1,\ldots,J$, but its pseudo p-value can still be small, if many random samples under the ideal strategy have even smaller SMD, i.e., $P(\Delta_j(g,h) \le \boldsymbol \Delta_j), j=1,\ldots, J,$ or $F^{\dagger}_\delta(\mathbf R^{\delta}-1)$ is high.  This is quite possible if the ideal strategy is highly restrictive or the study population is very homogeneous. Consider a silly example where the ideal strategy is defined as random samples whose SMD must be smaller than the the minimum of the observed $\boldsymbol \Delta_j, j=1,\ldots, J$. If there exists at least one sample by this ideal strategy, 
then the pseudo p-value is always 0. Vice versa, the SMD of the observed sample may appear to be large, but the pseudo p-value can still be large. This scenario can arise when the study population is highly heterogeneous so that many random samples under the ideal strategy have larger SMD than the observed sample. 

By Definition 1, the ideal sampling scheme should be able to provide samples with good CB.  Similar to \eqref{eqn:p}, we can define the random pseudo p-value for a random sample $g,h$, denoted as
\begin{equation}
 p(g, h) = \inf_\delta \big [ 1- F^\dagger_\delta (R^\delta_{g,h}-1) \big ] = 1 - \sup_\delta F^\dagger_\delta (R^\delta_{g,h}-1).     
\end{equation}
We shall use the distribution of $p(g,h)$ as a benchmark to define the standardized pseudo p-values later. 

We present some theoretical  properties of the pseudo p-value $\mathbf p$ and the random pseudo p-value $p(g,h)$ in three theorems below.  Proofs for Theorem 1-3  are outlined in the Appendix.  

\begin{flushleft}
\textbf{Theorem 1.} Some simple properties are as follows.
i) The pseudo p-value for the observed sample $M, N$ is $0 \le \mathbf p \le 1$. If $M, N$ can be generated by the ideal sampling strategy, then $0 < \mathbf p \le 1$.
ii) The random pseudo p-value $p(g, h)$ is a discrete random variable, and $0 < p(g,h) \le 1$. 
iii) If there exists $\delta$ such that $P(R^{\delta}_{g,h}\ge \mathbf R^{\delta} ) \le c$, then $\mathbf p \le c$. 
\end{flushleft}

In Theorem 1, all bounds for pseudo p-values  and discreteness of $R(g,h)$  follow directly from  definitions  above.  The following is a toy example to illustrate Theorem 1. 

\begin{flushleft}
\textit{A simple example for Theorem 1:} 
\end{flushleft}
Consider a small population $\mathcal U=\{1,2,3,4\}$ and the observed sample $M=\{1\}, N=\{2\}$.  The ideal sampling scheme is a cluster sampling where the population is split into two clusters, one cluster is chosen with equal probability, and then  the selected cluster is randomized to two arms. 
Assume that the two ideal clusters are $\{1, 2\}$  and   $\{3, 4\}$. There are a total of four distinct random samples $g,h$ under the ideal strategy but only two distinct $R^{\delta}(g,h)$ for any $\delta$ due to symmetry. The observed sample is one of the four ideal samples. 
If there exists $\delta$ such that the observed sample $\{1, 2\}$ has more imbalanced dimensions than the two ideal samples when the second cluster $\{3, 4\}$ is chosen, then $\mathbf p=.5$.  Otherwise, $\mathbf p=1$.  If instead the observed sample is $M=\{1\}, N=\{3\}$, then potentially $\mathbf p=0$ if there exists $\delta$ such that the observed sample $\{ 1, 3\}$ has more imbalanced dimensions than any of the four ideal samples. In any case, the support of $p(g,h)$ always consists of two points $( .5, 1)$. 

~\\

\begin{flushleft}
\textbf{Theorem 2.} (Right skewness of the random pseudo p-value)
If there exists $\delta$ such that $R^\delta_{g,h}$ is not degenerate, 
then 
\begin{displaymath}
    P(p(g,h) \le c) \ge 1 - \lceil 1-c \rceil,
\end{displaymath} where 
$1 - c\le \max_{a: F^\dagger_{\delta}(a)<1}F^\dagger_{\delta}(a)$,
and $\lceil 1-c \rceil$ is the smallest point on the partial support of $F^\dagger_{\delta}(R^{\delta}_{g,h})$ where 
$F^\dagger_{\delta}(R^\delta_{g,h}) > 1-c$, i.e.,
$\lceil 1-c \rceil
=\min\{b: P(F^\dagger_{\delta}(R^{\delta}_{g,h})=b)>0, b> 1-c\}$.
\end{flushleft}
~\\

To understand the implication of Theorem 2, consider the scenario where the population size is not too small and the ideal strategy is not overly stringent. Then conceivably there are many $\delta$ with non-degenerate $F^{\dagger}_{\delta}$.   Among the many $\delta$, some may give $\lceil 1-c \rceil \approx 1-c$, and thus $P(p(g,h) \le c) \gtrsim c$.  Roughly speaking, compared with a uniform random variable  $U(0,1)$, the random pseudo p-value $p(g,h)$ tends to be smaller.  

~\\
\begin{flushleft}
\textbf{Theorem 3.} (Relative size)
Let two pairs of random samples $(v,w)$, $(v^*, w^*)$ be drawn independently, $|v|=|v^*|, |w|=|w^*|$. The sampling schemes for $(v,w)$ and $(v^*, w^*)$ as well as the ideal strategy can all be different.  Then  
\begin{displaymath}
P(p(v,w) < p(v^*,w^*)) \ge \sup_\delta P(R^\delta_{v, w} > R^\delta_{v^*,w^*}). 
\end{displaymath}
\end{flushleft}
~\\

Heuristically, Theorem 3 means that a sample generated from a ``worse" strategy tends to have a smaller pseudo p-value compared with one from a ``better" way, where the merit of a study design is judged by the distribution of the number of imbalanced dimensions given a cutoff $\delta$. A poor strategy has a higher chance of yielding more imbalanced dimensions than a good strategy, and consequently a higher chance of having a smaller pseudo p-value. 
This property suggests that pseudo p-values can be used to compare samples. 

We know that pseudo p-values under the ideal scheme tend to be smaller than a random number from the uniform distribution. By Theorem 3, real samples which are often not from the ideal strategy may have even smaller $\mathbf p$. In many cases, $\mathbf p$ can be excessively small so that the conventional cutoff such as .05 may become overly stringent.  To tackle this potential problem, We define the standardized pseudo p-value $p^*$ using the distribution of random $p(g,h)$.

~\\
\begin{flushleft}
    \textbf{Definition 2.}  The standardized pseudo p-value is defined as 
    \begin{equation}\label{eqn:ps}
\mathbf p^* = P( p(g,h) \le \mathbf p),
\end{equation}
where $g, h$ is a random sample drawn by the ideal strategy.
\end{flushleft}

~\\

The standardized pseudo p-value $\mathbf p^*$ essentially represents the ranking of the observed sample among all random samples by the ideal strategy in pseudo p-values. For clarify, we report $\mathbf p^*$ in the percentage scale. Ignoring ties, it is a simple monotonic transformation from $\mathbf p$ to $\mathbf p^*$. 
It is reasonable to consider a small proportion of samples from the ideal strategy with the lowest pseudo p-values as relatively poor in CB.  By this rationale, we may use a cutoff such as 10\%, 20\%, or 25\% for the standardized pseudo p-value. For example, the observed sample is assessed as imbalanced if $\mathbf p^*<20\%$, i.e., among the bottom 20\% in terms of the pseudo p-value compared with random samples from the ideal strategy.

\subsection{Computing pseudo p-values}
The following theorem provides the basis for an algorithm to calculate the pseudo p-value.  Refer to Appendix for a sketch of proof. 

~\\
\begin{flushleft} 
\textbf{Theorem 4.} Let $\Gamma$ be the partial support of $\Delta_j(g,h), j=1,\ldots, J$, where $\Delta_j(g,h) \le \max_{1\le k\le J} \boldsymbol 
\Delta_k$.  If $M, N$ cannot be generated by the ideal strategy, $\Gamma$ further includes all  $\boldsymbol \Delta_k, k=1,\ldots, J$. Then 
    \begin{equation}\label{eqn:pc}
    \begin{split}
\mathbf p  = \min_{\delta \in \Gamma}  \big \{ 1- F^\dagger_\delta (\mathbf R^\delta-1) \big \} = 1 - \max_{\delta \in  \Gamma} F^\dagger_\delta (\mathbf R^\delta-1).
\end{split}
\end{equation}
\end{flushleft}

~\\

Due to the potentially large number of elements in $\Gamma$, to save computation cost in practice, we replace the theoretical $\Gamma$ in Theorem 4 by a fine grid of small positive numbers, e.g., $.01, .02, \ldots, 2.00$, where the maximum of the grid should be bigger than the maximum SMD in the observed sample. For each $\delta \in \Gamma$, $\mathbf R^\delta$ is known. The only computational task for \eqref{eqn:pc} is to calculate the CDF $F^\dagger_\delta$, which  should  be fully known under the finite population setting and given the ideal sampling scheme.  When the population size is small, we can simply calculate $F^\dagger_\delta$ by enumerating all random samples under the ideal scheme.  When the population size is moderately large, enumeration can be too expensive for computing. In these cases, we can estimate $F^\dagger_\delta$ by the Monte Carlo method, i.e.,  repeated samples by the ideal strategy.  The following algorithm presents further computational details, where we omit the estimation error assuming a large number of Monte Carlo iterations. 

~\\

\begin{flushleft}
    \textit{An algorithm to compute $\mathbf p$ and $\mathbf p^*$:}
    \begin{enumerate}
        \item Preparations: nominate an ideal sampling strategy;  set the search grid $\Gamma$; prepare the $K \times J$ input data matrix $( \mathbf X_1', \mathbf X_2',\ldots, \mathbf X_K')'$; and specify the observed sample $M, N$. Continue to step 2 or 3. 
        \item Enumeration: if $K, |M|, |N|$ are reasonably small, enumerate all possible samples under the ideal strategy using the sample size $|M|, |N|$. 
        \item Monte Carlo: instead of enumeration, repeatedly draw random samples from $\mathcal U$ by the ideal strategy, e.g., 10,000 rounds. 
        \item Repeat the next steps 5 to 7 for each $\delta \in \Gamma$
        \item For each random sample $g, h$ generated by step 2 or step 3, count $R^{\delta}_{g,h}$
        \item Calculate the CDF $F^\dagger_{\delta}$ using the full population or the Monte Carlo sample of $R^{\delta}_{g,h}$.
        \item Count $\mathbf R^{\delta}$ for the observed sample $M, N$; using $F^{\dagger}_{\delta}$ given in the last step, calculate $F^{\dagger}_{\delta} (\mathbf R^{\delta}-1)$.
        \item Calculate $\mathbf p$ by \eqref{eqn:pc} using the collection of results $F^{\dagger}_{\delta} (\mathbf R^{\delta}-1), \delta \in \Gamma$. 
        \item Repeat step 8 treating each random sample from step 2 or 3 as the observed sample. This gives the full population or the Monte Carlo sample of $p(g,h)$.
        \item Calculate $\mathbf p^*$ by its definition \eqref{eqn:ps}.
    \end{enumerate}
\end{flushleft}
~\\

This algorithm consists of a relatively large number of iterations, but each iteration is simple counting for an enumerated or a Monte Carlo sample. Calculating $F^{\dagger}_\delta, \delta\in \Gamma$ is straightforward. The computational efficiency of this algorithm can be high in a computing environment optimized for vector and matrix operations.

\section{Simulations}
\subsection{Simulation settings}
We conduct a simulation study to numerically examine the performance of the pseudo p-value and the standardized pseudo p-value. The overall simulation strategy is as follows. We first generate a finite population under a certain setting of all known constants, called a scenario. Next, we draw a sample by a study design we nominate, where our study designs mimic typical evaluation study designs in practice. The pseudo p-value and the standardized p-value for each sample are calculated.  We repeat this process 1,000 times under each simulation scenario. Finally, we aggregate results across the 1,000 iterations by graphical summaries and descriptive statistics to examine the pseudo p-value's performance in assessing CB for different study designs. Details for each step are provided below. 

A scenario consists of the population $\mathcal U$, the characteristics $\mathbf X_i, i=1,\ldots, K$, and the sample size $|M|, |N|$.  The number of covariates is $J=10$ in all scenarios. 
We consider five population sizes: $K=$12, 16, 20, 100, and 400.  The characteristics of the first half of $\mathcal U$ are  generated by $\text{i.i.d. } N(0,1)$ and those of the second half are generated by $\text{i.i.d. } N(bias,1)$, where the bias term is between 0 and 2. 
Excessively large bias terms for some $K$ values are omitted due to obvious results in scenarios with the same $K$ but smaller biases.  The sample size $|M|, |N|$ is set between 10\% and 20\% of $K$ in all scenarios. 
There are a total of 16 scenarios with a small population size in which we apply the enumeration method to calculate the pseudo p-value, and 8 scenarios with a relatively large populations size where we use 10,000 Monte Carlo samples to calculate each pseudo p-value. 

We consider six study designs, i.e., sampling strategies, to draw $M, N$, which are abbreviated as \textit{Randomized}, \textit{Segregated}, \textit{Partial}, \textit{Matched}, \textit{R\_Partial}, and \textit{Natural}.  All study designs are sampling without replacement so that $M$ and $N$ do not overlap. 
\begin{itemize}
\item \textit{Randomized} is SRS and the ideal strategy for calculating all pseudo p-values.  
\item 
\textit{Segregated} draws a random sample $N$ from the first half of $\mathcal U$ and a random sample $M$ from the second half of $\mathcal U$, which mimics a systematically biased observational study design with varying levels of biases across simulation scenarios.  

\item \textit{Partial} draws $N$ and a fixed part of $M$ from the first half of $\mathcal U$, and then draws the remainder of $M$ from the second half of $\mathcal U$.  Specifically, scenarios 1 to 12 under the enumeration method have 1 EU of $M$ drawn from the first half of $\mathcal U$; scenarios 13 to 16 under the enumeration method have 2 EU of $M$ drawn from the first half of $\mathcal U$; and all scenarios under the Monte Carlo method set 8 EU of $M$ drawn from the first half of $\mathcal U$. This design mimics an imperfect matching operation to remove imbalances in observed characteristics or a restricted natural experiment that does not have the full freedom to choose control EU.

\item \textit{Matched} draws both $N$ and $M$ from the first half of $\mathcal U$. This design serves the role of a good matching operation to remove imbalances in observed characteristics.

\item \textit{R\_Partial} is a variant of \textit{Partial}, where the part of $M$ drawn from the first half is a random number centered at $.5|M|$. 
\item \textit{Natural} draws $N$ from the first half of $\mathcal U$, and then draws $M$ by SRS. This design mocks a natural experiment where the treated arm is 
self-selected and the control arm is representative of the study population. 
\end{itemize}

\subsection{Simulation results}
Due to the large volume of results, we elect to present the results of the 8 scenarios using the Monte Carlo method.  The results of the 16 scenarios using the enumeration methods are included in supplemental materials. 
The empirical distributions of the 1,000 instances of $\mathbf p$ and $\mathbf p^*$ are summarized by box plots in Figure 1, where the left frame illustrates $\mathbf p$ and the right frame does $\mathbf p^*$.  
In the left frame of Figure 1, the six study designs do not differ when the bias term is zero (scenarios 1 and 6), all of which show a skewed distribution of $\mathbf p$ to the right.  
When the bias term is non-zero, the ideal strategy \textit{Randomized} and \textit{Matched} do not have notable changes. By contrast, the less-than-ideal study designs (\textit{Segregated}, \textit{Partial}, \textit{R\_Partial}, 
\textit{Natural}) all become more skewed to the right, where the skewness is extreme when the bias term is large (scenarios 5 and 8). 
Although rigorously speaking these box plots do not represent the random pseudo p-value $p(g,h)$, these observations still shed some light on Theorems 2 and 3. 

In the right frame of Figure 1, it is clear that the distribution of the standardized pseudo p-value is generally less peaked and more centered toward .5, compared with the distribution of $\mathbf p$. This makes it easier to visually discern differences among study designs. 
  However, the essential findings regarding comparing study designs remain unchanged, due to the monotonic transformation between $\mathbf p$ and $\mathbf p^*$, 

Table 3 reports the sample proportions that a particular study design has the smallest $\mathbf p$. This proportion can be interpreted as the chance for a study design to be assessed as the best among all six competitors in terms of the goodness of CB.  When the bias term is 0, all designs have a roughly equal chance to be the best.  When the bias term is not zero, \textit{Matched} and \textit{Randomized} are  most likely to be the best, while the difference between the two is small.  The four less-than-ideal study designs still have a small chance to be assessed as the best: \textit{R\_Partial} and \textit{Natural} have higher chances than \textit{Partial}, and the chance for \textit{Segregated} to win the competition is the lowest. These results demonstrate that the pseudo p-value can yield adequate recommendations for comparing alternative sampling strategies.

Tables 4 and 5 report the sample proportions that a particular study design yields a sample with an unacceptable level of CB based on the empirical threshold we have introduced earlier, i.e., .05 for $\mathbf p$ and $20\%$ for $\mathbf p^*$. Coincidentally, by either cutoff, \textit{Randomized} always has about 20\% chance to be declared as unacceptable.  The chance  for \textit{Matched} to be unacceptable is similar and sometimes slightly lower than that of \textit{Randomized}.  When the bias term is non-zero, the less-than-ideal study  designs all have greater chance to be unacceptable than the two good designs.  Similar to Table 3,  \textit{R\_Partial} and \textit{Natural} have a lower chance to be unacceptable, compared with \textit{Partial}.  The chance to declare \textit{Segregated} as unacceptable can be very high when the bias is large. These results demonstrate that the empirical thresholds on pseudo p-values result in reasonable decisions to assess a study design.

\section{Assess CB in the California SSB Tax Study}
The California SSB Tax Study is an ongoing policy evaluation study using a natural experiment design. The intervention arm consists of four California cities that have adopted the SSB excise tax between 2015 and 2017 (Albany, Berkeley, San Francisco, and Oakland). The four intervention cities are all located in the greater San Francisco Bay area with distinct city-level characteristics.  Potential control cities include 328  incorporated cities in California, where health care is covered by KP and no SSB excise tax has been in effect between 2009 and 2020. The study population of EU consists of a total of $K=332$ cities in California. 

Seventeen city-level covariates are collected: total population, population density, \% population having KP membership, \% males, \% females, \% population in each of the four age strata ($\le 19$, 20-44, 45-64, $\ge 65$),  \% population in each of the four race/ethnicity categories (Latino of all races, non-Latino African American, non-Latino white, non-Latino Asian and others), \% population living below poverty line, \% population in each of the three education attainment levels (high school diploma/GED or lower, some college or associate degree,  bachelor degree or higher). 
Except for \% population having Kaiser Permanente membership, all other covariates are public information using 5-year averages of U.S. Census American Community Survey prior to SSB tax implementation years. 
Weighted Euclidean distance metrics are calculated using 13 or the 17 covariates by removing a redundant category from factors (e.g., dropping \% female from the factor of sex) and after standardization. 
Weights are assigned to down weight percentages in the same factor (e.g., the three percentages for age strata each had a weight of .333) so that the total weight of a factor with more than one level is 1.  Each intervention city is matched to 10 control cities with generally the shortest distance metric. Some ad hoc adjustments are made to avoid overlap in selected control cities among different intervention cities and to ensure no control cities border any intervention cities.  Supplemental materials provide the list of the 44 study cities and all SMD in the study sample.  As discussed in the introduction, evaluating the goodness of CB in the selected study sample is essential for the validity of future study findings. 

We calculate the pseudo p-value and the standardized pseudo p-value to assess the level of CB of all 17 city-level covariates in the selected study sample.  Following the algorithm in Section 2.4, we prepare the input data $\mathbf X_i, i=1,\ldots 332$, where $\mathbf X_i$ consists of all 17 city-level covariates.  We set $|M|=4, |N|=40$ to calculate the pseudo p-values for the overall study sample and $|M|=1, |N|=10$ to assess CB for each subsample of one intervention city and its 10 matched control cities. We use 
$\Gamma=c(.01,.02,\ldots,3.00)$ Which covers all observed SMD values (cf. Theorem 4). We use SRS as the ideal sampling strategy in all calculations.  Due to the relatively large population and sample size, we use 100,000 Monte Carlo rounds to calculate pseudo p-values.  

Table \ref{tb:case} reports the results. The overall sample has $\mathbf p=.132$ and $\mathbf p^*=40.4\%$, both are above the recommended cutoff in this paper. For subsamples by individual intervention cities, $\mathbf p$ ranges between .15 and .37, and $\mathbf p^*$ is between 40\% and 72\%.   Therefore, we may conclude that in relation to an ideal simple randomization, both the overall study sample and the four subsamples have acceptable levels of CB.  
However, those pseudo p-values suggest that the goodness of CB of the study sample, which is nearly the best match we can find, is only about average among all SRS samples. 
It is also worth noting that, individual SMD $\boldsymbol \Delta_j$ can be large (see Table \ref{tb:case} and more details in supplemental materials). 
These facts are likely due to the peculiarity of the four treated cities: no other study cities  can be an absolutely perfect match to the treated cities. 
However, due to the high level of heterogeneity among the entire study population, an ideal SRS yields many samples with worse CB than our selected study sample.

Lastly, we note that the California SSB Tax Study exemplifies the excessive restrictions by the ad hoc procedure reviewed in Section 2.2. For the full sample setting $|M|=4, |N|=40$ and if we use a conventional cutoff of $\delta=.1, r=2$, i.e., requiring no more than 2 dimensions with SMD greater than .10, none of the 100,000 SRS samples can satisfy these cutoffs. 
Increasing cutoffs to $\delta=.2, r=2$ yields only 48 SRS samples as balanced. Further increasing cutoffs to $\delta=.3, r=2$ yields 1.35\% SRS samples assessed as balanced.  We can have 7.81\% SRS samples assessed as balanced if we are willing to accept 4 imbalanced dimensions $\delta=.30, r=4$.  
 Although some searching algorithm can still find these few samples, this would be a highly constrained study design with questionable generalizability.  Similarly, the situation is even poorer for the subsample setting $|M|=1, |N|=10$: only 23 of the 100,000 SRS samples can be assessed as balanced using  the most lenient cutoffs above ($\delta=.3,  r=4$). 
This real-data example further strengthens the message from the crude normal-binomial approximation in Table 1.

\section{Discussion}
The selection of the EU in the study sample is foundational for an evaluation study, as important as the numerous technical aspects in the data analysis. 
Assessing the goodness of CB of the selected sample is a usual practice in many evaluating studies.
In our California SSB Tax Study, the treated arm consists of four cities, all having very distinctive characteristics in their health care and population dimensions, some possessing a national or international recognition for their uniqueness. Selecting control cities is crucial to ensure an unbiased evaluating study, findings from which can have important implications for informing future tax and food policies regarding added sugar in commercial food supplies.  It is necessary to apply a matching procedure to find control cities with a good level of CB compared with the treated cities. 
However, in a finite population and with a relatively small sample size, the usual ad hoc procedure to assess the goodness of CB is overly restrictive or nearly impossible to apply.  Based on the finite population inference framework, we have proposed the novel concept of the pseudo p-value and its minor variant, the standardized pseudo p-value.  These two pseudo p-values are simple descriptive statistics with convenient implementation and interpretation.  
We have conducted numerical simulation studies to demonstrate their use. 
The use of pseudo p-values in the California SSB Tax Study reveals the strength and weakness of the selected study sample.

The pseudo p-values are unrelated to the classic hypothesis test but share some superficial features. First and foremost, the pseudo p-value is a tail probability under a presumed distribution, which is similar to a classic p-value.  The key concept of $\mathbf R^\delta$ is analogous to both the false positives in multiple hypothesis tests and the test statistic in a single hypothesis test. The algorithm to calculate pseudo p-values is somewhat similar to resampling-based inference methods such as the permutation test. 

The theoretical properties of pseudo p-values  in this paper have much room to be strengthened or extended. However, these rudimentary results suffice to provide the most important features a CB assessment measure should have: to assess whether an observed sample has an acceptable level of CB and to compare among alternative samples to find the one with the best CB. As an application-driven study, we do not delve into deeper mathematical endeavors in this paper.

There are many possible extensions to the pseudo p-values in this paper. As aforementioned, a straightforward extension is to replace the SMD by other measures suitable for discrete variables, such as Cohen's h or odds ratio. Descriptive statistics for more than two arms, e.g., between-arm variance and the F statistics from one-way ANOVA, can be used to accommodate studies with more than two arms.  It is also possible to further extend to studies with a continuous treatment dosage, where correlation or $R^2$ can serve the role of SMD in two-arm settings. 
Another possible extension is to allow the observed sample $M, N$ be random instead of observed. We can then define the mean of the random pseudo p-value as an CB measure. This extension will make the pseudo p-value applicable to propensity score weighting studies where the weights can be interpreted as a sampling strategy. 
The key quantity underlying the pseudo p-value, $\mathbf R^\delta$, can be replaced by a weighted count to reflect the relative importance of covariates in a specific scientific context.  We plan to study these extensions in future work.

\section*{Appendix}
\subsection*{Sketch of proofs}
We outline proofs of Theorems 1 to 4 in this appendix. First, Theorem 1 follows directly from the definitions. In i) note that when $M, N$ can be generated by the ideal strategy, there is at least some non-zero probability to observe the same number of imbalanced dimensions so that $\mathbf p>0$. The discreteness of $p(g,h)$ is due to the fact that in the finite population setting, the underlying probability space that defines all distributions in this paper is discrete. 

Proof of Theorem 2 is based on the following lemma. 

{\flushleft \textbf{Lemma.} }
Let $X$  be a non-degenerate and discrete random variable.  The support of $X$ consists of a finite set of natural numbers and $X$ has a CDF $F_X$. Then $P(F_X(X-1) \le a) = \lceil a \rceil $, where 
$0  \le a \le \max_{F_X(x)<1}F_X(x)$, and $\lceil a \rceil$ is the smallest point on the partial support of $F_X(X)$ where  
$F_X(X) > a$, i.e.,
$\lceil a \rceil
=\min\{b: P(F_X(X)=b)>0, b > a\}$.

\begin{flushleft}
    \textit{Proof of Lemma.}
\end{flushleft}

Write the support of $X$ as $\{x_i, i=1,\ldots, s\}$ with $P(X=x_i)=p_i>0$, $s\ge 2$, and all $x_i$ are natural numbers. 
Let $c_0=0$, $c_i = \sum_{j\le i} p_i, i=1,\ldots,s \}$.  
The support of $F_X(X)$ is $\{ c_i , i=1,\ldots,s \}$ with corresponding probabilities $P( F_X(X) = c_i) = P(X=x_i) = p_i$. 
The support of $F_X(X-1)$ is $\{ c_i, i=0,\ldots,s-1 \}$ with corresponding probabilities $P( F_X(X-1) = c_i) = P(F_X(X) = c_{i+1}) = p_i$.

By definition, $\lceil a \rceil \in \{ c_i, i=1,\ldots, s \}$. Denote $c_r = \lceil a \rceil, r\ge 1$. 
By this system of notation, 
we have
 $0 \le c_{r-1} \le  a < c_r \le 1$. 
It is 
\begin{displaymath}
\begin{split}
&P(F_X(X-1) \le a) 
= P(F_X(X-1) \le c_{r-1})
= P( F_X(X) \le c_r )
= \sum_{j\le r} p_j 
= c_r = \lceil a \rceil. \square
\end{split}
\end{displaymath} 

\begin{flushleft}
    \textit{Proof of Theorem 2.}
\end{flushleft}

Apply Lemma to  $R^{\delta}_{g,h}$,
\begin{displaymath}\begin{split}
&P(p(g,h)\le c) 
= P(\inf_\delta \big \{ 1- F^\dagger_\delta (R^{\delta}_{g,h}-1) \big \}  \le c ) 
\ge 
1 - P(F^\dagger_{\delta} (R^{\delta}_{g,h}-1) \le 1-c) 
=1-\lceil 1-c \rceil. ~~\square
\end{split}
\end{displaymath}

\begin{flushleft}
~\\
\textit{Proof of Theorem 3.}  
\end{flushleft}
\begin{displaymath}
\begin{split}
&P(p(v,w)-p(v^*,w^*)<0)  \\=&
 P( 1 - \sup_\theta F^\dagger_\theta (R^{\theta}_{v,w}-1)
  - 1 +\sup_\delta F^\dagger_\delta (R^{\delta}_{v^*,w^*}-1) < 0)\\
\ge& P\big ( \sup_\delta[  F^\dagger_\delta (R^\delta_{v^*,w^*}-1)
 -  F^\dagger_\delta (R^\delta_{v,w}-1)] <0 \big  ) \\
=&  \sup_\delta P \big ( R^\delta_{v^*,w^*}-
 R^\delta_{v,w}  < 0 \big ). \square
\end{split}
\end{displaymath}

\begin{flushleft}
\textit{Proof of Theorem 4.}  
\end{flushleft}

First, for $\delta$ out of the range of $\Gamma$, $F^{\dagger}_\delta(\mathbf R^\delta)$ is always 0 and so does not play a role for $\mathbf p$.  
Next, consider a left-close and right-open interval, denoted by $\kappa$, between two adjacent and sorted elements with distinct values in $\Gamma$.  By construction,  $\mathbf  R^{\delta}$ is a constant for all $\delta \in \kappa $. 
For any $\delta_1, \delta_2 \in \kappa$,  $R^{\delta_1}_{g,h}=R^{\delta_2}_{g,h}$, w.p. 1. Therefore, $F^{\dagger}_{\delta_1}(\mathbf  R^{\delta_1}-1)=F^{\dagger}_{\delta_2}(\mathbf  R^{\delta_2}-1)$ for $\delta_1, \delta_2 \in \kappa$.  This implies that  
$F^{\dagger}_\delta(\mathbf R^{\delta} -1 )$ is a step function of $\delta$, whose value can jump only at elements of $\Gamma$. 
$\square$

\newpage 
\linespread{1}

\begin{table}[h]\label{tb:1}
\center 
\caption{Approximate probabilities of accepting SRS samples $g, h$ as balanced. $P( I\{ \Delta_{j,g,h} \ge \delta\}$ is the probability for declaring a selected dimension as imbalanced. $P(R^\delta_{g,h}\le 1|J)$ is the probability of declaring samples $g,h$ as balanced given the total number of dimensions $J$.}
\begin{tabular}{c|c|c  |r|r|r}
\toprule
    $\delta$ & $|g|$ &$|h|$  & $P( I\{ \Delta_j(g,h) \ge \delta\} )$ &$P(R^\delta_{g,h}\le 1|J=10)$ &$P(R_{g,h}^\delta \le 2|J=20)$ \\
      \hline  
.2&4& 40 & .703 & .0001 & 3.2e-8 \\
.3&4& 40 & .567 & .003 & 1.9e-5 \\
.2&10&10 &.654 &.0005 & 4.2e-7 \\
.3&10&10 &.502 &.0103 & 1.8e-4 \\
.2&40&40 &.371 &.067 & .007 \\
.3&40&40 &.178 &.440 & .276 \\
.2&100&100&.157&.518 & .370\\
.3&100&100&.034&.957 & .971\\
\bottomrule 
\end{tabular}
\end{table}

\begin{table}[ht]
    \centering
    \caption{Data generating scenarios in the simulation study.}
    \label{tab:data_plan}
    \begin{tabular}{ccccc|ccccc}
    \toprule 
    \multicolumn{5}{c|}{Enumeration}&\multicolumn{5}{c}{Monte Carlo}\\
    Scenarios &  $K$ & $|M|$ & $|N|$ & Bias& 
    Scenarios &  $K$ & $|M|$ & $|N|$ & Bias\\
    \hline
    1 & 12 & 2 & 2 & 0 & 1 & 100 & 20 & 20 & 0 \\
    2 & 12 & 2 & 2 & .5 & 2 & 100 & 20 & 20 & .1 \\
    3 & 12 & 2 & 2 & 1 & 3 & 100 & 20 & 20 & .25 \\
    4 & 12 & 2 & 2 & 2 & 4 & 100 & 20 & 20 & .5 \\
    5 & 16 & 2 & 2 & 0 & 5 & 100 & 20 & 20 & .75 \\
    6 & 16 & 2 & 2 & .5 & 6 & 400 & 40 & 40 & 0 \\
    7 & 16 & 2 & 2 & 1 & 7 & 400 & 40 & 40 & .1 \\
    8 & 16 & 2 & 2 & 2 & 8 & 400 & 40 & 40 & .25 \\
    9 & 20 & 3 & 3 & 0 & & & & & \\
    10 & 20 & 3 & 3 & .5 & & & & & \\
    11 & 20 & 3 & 3 & 1 & & & & & \\
    12 & 20 & 3 & 3 & 2 & & & & & \\
    13 & 20 & 4 & 4 & 0 & & & & & \\
    14 & 20 & 4 & 4 & .5 & & & & & \\
    15 & 20 & 4 & 4 & 1 & & & & & \\
    16 & 20 & 4 & 4 & 2 & & & & & \\
    \bottomrule 
    \end{tabular}
\end{table}

\begin{table}[ht]
    \centering
    \caption{Sample proportions among 1,000 iterations that a study design has the largest $\mathbf p$.}
    \begin{tabular}{ccc|cccccc}
    \toprule
     Scenario & $K$ & Bias & \textit{Randomized} & \textit{Segregated} & \textit{Partial} & \textit{Matched} & \textit{R\_Partial} & \textit{Natural}  \\
     \hline
    1 & 100 & 0 & .17 & .15 & .18 & .17 & .17 & .17 \\
    2 & 100 & .10 & .19 & .11 & .18 & .18 & .17 & .17 \\
    3 & 100 & .25 & .28 & .04 & .15 & .23 & .15 & .14 \\
    4 & 100 & .50 & .38 & .00 & .05 & .39 & .12 & .06 \\
    5 & 100 & .75 & .41 & .00 & .01 & .52 & .04 & .01 \\
    6 & 400 & 0 & .18 & .18 & .17 & .16 & .14 & .16 \\
    7 & 400 & .10 & .22 & .11 & .17 & .20 & .16 & .14 \\
    8 & 400 & .25 & .33 & .02 & .05 & .33 & .15 & .12 \\
    \bottomrule
    \end{tabular}
\end{table}

\begin{table}[ht]
    \centering
    \caption{Sample proportions  among 1,000 iterations that a study design has $\mathbf p<.05$.}
    \begin{tabular}{ccc|cccccc}
    \toprule
     Scenario & $K$ & Bias & \textit{Randomized} & \textit{Segregated} & \textit{Partial} & \textit{Matched} & \textit{R\_Partial} & \textit{Natural}  \\
     \hline
1	&	100	&	0	&	0.22	&	0.22	&	0.22	&	0.23	&	0.23	&	0.22	\\
2	&	100	&	0.1	&	0.22	&	0.28	&	0.26	&	0.23	&	0.23	&	0.25	\\
3	&	100	&	0.25	&	0.22	&	0.6	&	0.37	&	0.23	&	0.31	&	0.38	\\
4	&	100	&	0.5	&	0.21	&	0.99	&	0.71	&	0.19	&	0.55	&	0.73	\\
5	&	100	&	0.75	&	0.20	&	1	&	0.95	&	0.15	&	0.80	&	0.93	\\
6	&	400	&	0	&	0.21	&	0.19	&	0.18	&	0.19	&	0.21	&	0.21	\\
7	&	400	&	0.1	&	0.21	&	0.32	&	0.27	&	0.18	&	0.25	&	0.25	\\
8	&	400	&	0.25	&	0.20	&	0.83	&	0.64	&	0.18	&	0.39	&	0.47	\\
    \bottomrule
    \end{tabular}
\end{table}

\begin{table}[ht]
    \centering
    \caption{Sample proportions  among 1,000 iterations that a study design has $\mathbf p^*<20\%$.}
    \begin{tabular}{ccc|cccccc}
    \toprule
     Scenario & $K$ & Bias & \textit{Randomized} & \textit{Segregated} & \textit{Partial} & \textit{Matched} & \textit{R\_Partial} & \textit{Natural}  \\
     \hline
1	&	100	&	0	&	0.20	&	0.21	&	0.21	&	0.22	&	0.22	&	0.20	\\
2	&	100	&	0.1	&	0.21	&	0.26	&	0.25	&	0.21	&	0.21	&	0.24	\\
3	&	100	&	0.25	&	0.20	&	0.58	&	0.34	&	0.21	&	0.29	&	0.36	\\
4	&	100	&	0.5	&	0.20	&	0.99	&	0.69	&	0.18	&	0.53	&	0.71	\\
5	&	100	&	0.75	&	0.19	&	1	&	0.95	&	0.14	&	0.79	&	0.93	\\
6	&	400	&	0	&	0.21	&	0.19	&	0.18	&	0.19	&	0.20	&	0.21	\\
7	&	400	&	0.1	&	0.21	&	0.32	&	0.27	&	0.18	&	0.24	&	0.24	\\
8	&	400	&	0.25	&	0.20	&	0.82	&	0.64	&	0.17	&	0.39	&	0.46	\\
\bottomrule    \end{tabular}
\end{table}

\begin{table}[ht]
    \centering \label{tb:case}
    \caption{Pseudo p-values $\mathbf p$, standardized pseudo p-values $\mathbf p^*$, and descriptive statistics of observed $\boldsymbol \Delta_j, j=1,\ldots, 17$ for the four subsamples and the overall sample in the California SSB Tax Study.}
    \begin{tabular}{r|ccccccc}
        \toprule
        City & $\mathbf p$ &$\mathbf p^*$&\multicolumn{5}{c}{$ \boldsymbol \Delta_j$}  \\
        & & &Min &Q1 &Median&Q3&Max \\
        \hline
        Albany & .337 & 71.4\% & .075&.203& .369&.546&1.159\\
        Berkeley & .366 & 74.4\%& .077&.154&.657&.716&1.219 \\
        Oakland & .202 & 52.7\% &.033&.223&.478&.852&2.634 \\
         San Francisco & .159 & 46.3\% &
        .134 & .352& .462& .935 & 2.033\\
        Overall & .132 & 40.4\% &.018 &  .084&  .411& .790 & 1.142 \\
        \bottomrule
    \end{tabular}
\end{table}

\begin{landscape}
\begin{figure}
\centering
\includegraphics[width=10cm]{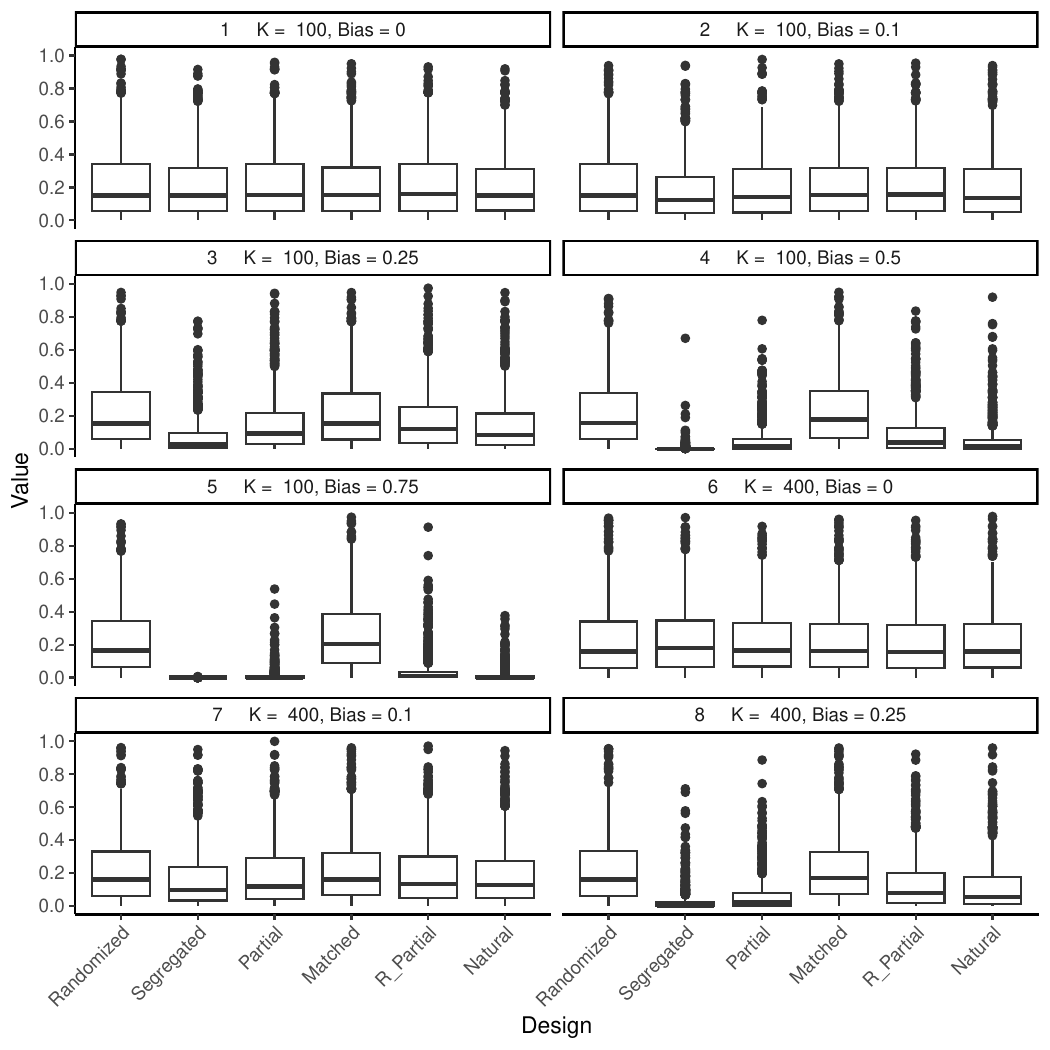}
\includegraphics[width=10cm]{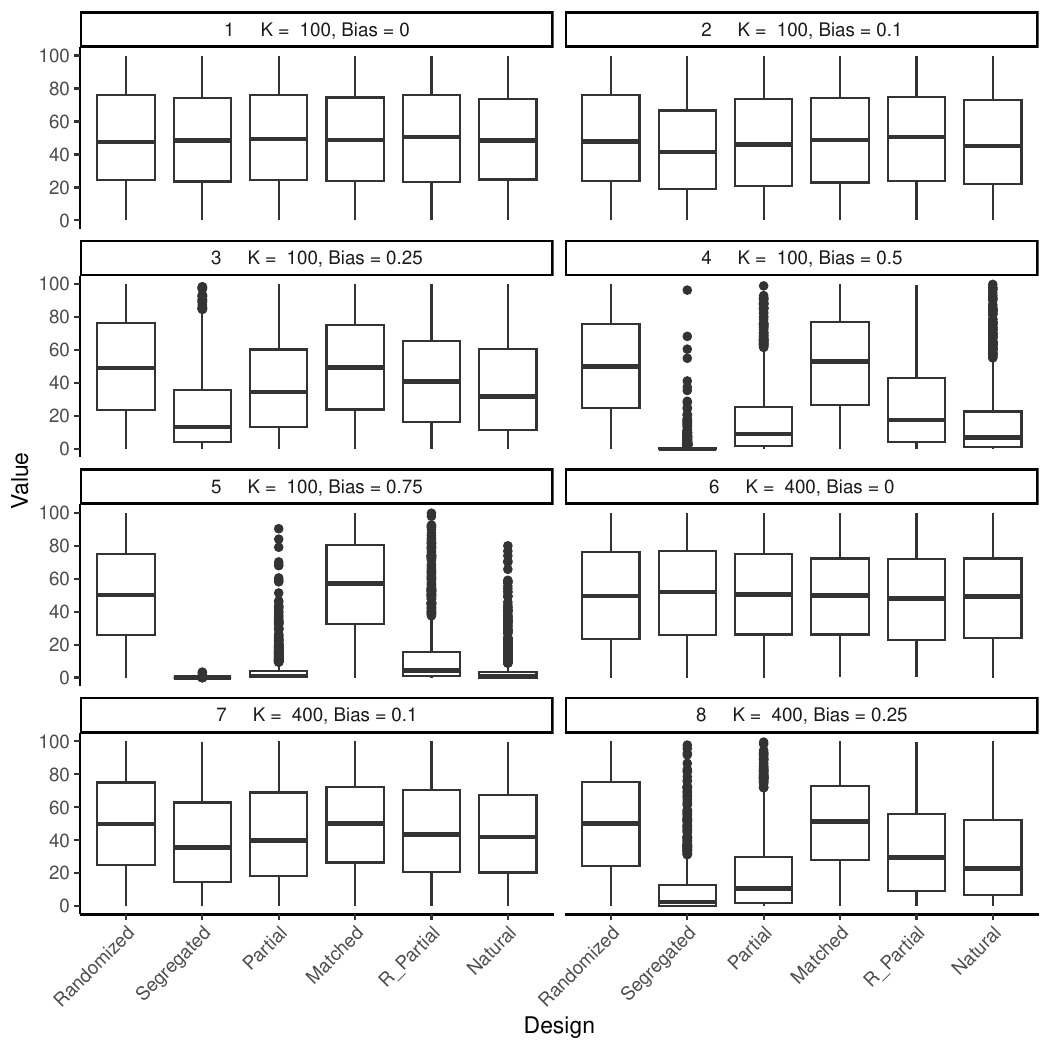}
\caption{Box plots of pseudo p-values $\mathbf p$ (left) and standardized pseudo p-values $\mathbf p^*$ (right) computed by the Monte Carlo
method in the 1,000 iterations: sub-frames are simulations scenarios and bars in each sub-frame correspond to the six study designs.}
\end{figure}
\end{landscape}

\begin{figure}[ht]
    \centering
\includegraphics[width=12cm]{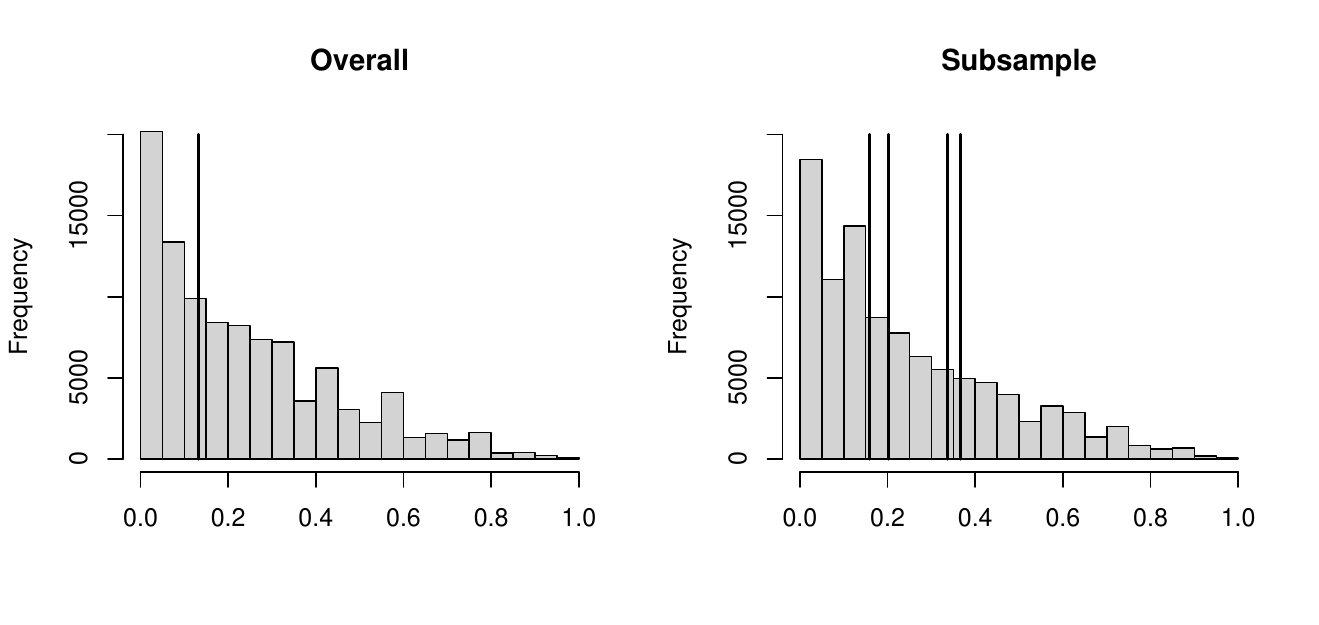}
    \caption{Histograms of the 100,000 random pseudo p-values $p(g,h)$  overlaid with $\mathbf p$ of the observed sample  in the California SSB Tax Study: (left) the overall sample with $|M|=4, |N|=40$ and the long vertical line marks $\mathbf p=.133$; (right) the subsample with $|M|=1, |N|=10$, where the vertical lines  from left to right are San Francisco, Oakland, Albany, and Berkeley.} 
\end{figure}

\newpage

\bibliography{bib}
\newpage

\end{document}